%% file: Bagnulo.tex
\documentclass[useAMS,usenatbib]{mn2e}
\usepackage{graphicx}
\newcommand{\rf}{\ensuremath{r}}
\newcommand{\pq}{\ensuremath{P_Q}}

\newcommand{\pqn}{\ensuremath{p_q}}
\title[Linear spectro-polarimetry of asteroids]{Linear spectro-polarimetry:
a new diagnostic tool for the classification and characterisation of asteroids}
\author[S. Bagnulo et al.]{S. Bagnulo$^{1}$, A. Cellino$^{2}$ and M. F. Sterzik$^{3}$\\
$^{1}$Armagh Observatory, College Hill, Armagh BT61 9DG, UK. {\rm E-mail: sba@arm.ac.uk}\\
$^{2}$INAF - Osservatorio Astrofisico di Torino, I-10025 Pino Torinese, Italy.
{\rm E-mail: cellino@oato.inaf.it}\\
$^{3}$European Southern Observatory, Karl-Schwarzschild-Strasse 2, D-85748 Garching, Germany. 
{\rm E-mail: msterzik@eso.org}
}

\begin{document}

\date{Accepted 2014 September 14. Received 2014 September 08; in original form 2014 July 24}

\pagerange{\pageref{firstpage}--\pageref{lastpage}} \pubyear{2014}

\maketitle

\label{firstpage}

\begin{abstract}
We explore the use of spectro-polarimetry as a remote
sensing tool for asteroids in addition to traditional reflectance
measurements. In particular we are interested in possible
relationships between the wavelength-dependent variation of linear
polarization and the properties of the surfaces, including albedo and
composition.
We have obtained optical spectro-polarimetric
measurements of a dozen asteroids of different albedo and
taxonomic classes and of two small regions at the limb of the
Moon.

We found that objects with marginally different relative reflectance
spectra (in the optical) may have totally different polarization
spectra. This suggests that spectro-polarimetry may be used to
refine the classification of asteroids.
We also found that in some cases the Umov law may be violated, that
is, in contrast to what is expected from basic physical
considerations, the fraction of linear polarization and the
reflectance may be positively correlated.  In agreement with a
few previous studies based on multi-colour broadband polarimetry, we
found that the variation of linear polarization with wavelength and
with phase-angle is correlated with the albedo and taxonomic class of
the objects.  Finally, we have serendipitously discovered that
spinel-rich asteroid (599)\,Luisa, located very close to the Watsonia
family, is a member of the rare class of Barbarian asteroids.

We suggest that future modelling attempts of the surface structure of
asteroids should be aimed at explaining both reflectance and
polarization spectra.
\end{abstract}


\section{Introduction}\label{Sect_Intro}
Light scattered by surfaces is polarized. This may be intuitively
understood by thinking that an electron sitting in a planar surface
and hit by an electromagnetic wave is more free to oscillate in the
direction parallel to the surface itself rather than perpendicular to
it. Accordingly, the radiation re-emitted by the electron is partially
linearly polarized in the direction parallel to the surface and
perpendicular to the scattering plane ({\it i.e.}, the plane
containing the incident and the scattered light beams).  Since the
radiation produced by the oscillations of an electron moving up and
down through the surface is more efficiently damped by a darker
surface than by a brighter one, one can expect that the light
reflected by a darker surface is more polarized than the light
reflected by a brighter surface. The state of the polarization of the
scattered radiation depends on the structure and composition of the
reflecting surface and on the scattering angle, and its measurement
may reveal information about the physical properties of the reflecting
surface.

Broadband linear polarization (BBLP) measurements have long been used
as a remote sensing tool for the characterisation of the objects of
our solar system. 
BBLP measurements in the standard optical filters are usually plotted
as a function of the phase-angle (the angle between the sun and the
observer as seen from the target object) and the morphology of the
resulting \emph{phase-polarization curves} may be used for the
purposes of albedo determinations \citep[see][and references
  therein]{Celetal12}, and for asteroid classification
\citep{Penetal05}.  Since main-belt asteroids orbit at a significantly
longer distance from the Sun than Earth, the phase-angles at which
they may be observed are restricted to a small interval, typically
$\sim 0 - 30\degr$. In the case of near-Earth objects the
maximum attainable phase-angle can be higher, well above
$40\degr$. Perhaps the most surprising feature of asteroid
polarimetric properties is that at small phase-angles the plane of
linear polarization is parallel to the scattering plane, in contrast
to the simple scattering mechanism sketched out above. This phenomenon,
which is traditionally referred to with the somehow confusing term of
\textit{negative polarization}, is normally seen in the
$0\degr-20\degr$ phase-angle range (usually referred to as the {\it
  negative branch} of the phase-polarization curve) and may be
explained in terms of coherent backscattering \citep{KarriAst3}.

A widely adopted remote-sensing tool for the physical
characterization of small solar system bodies is
spectroscopy. Similarly to what happens in stellar spectroscopy,
asteroid reflectance spectra are classified into distinct taxonomic
classes.  Taxonomy based on multi-band optical photometry was first
developed by several authors in the '70s, and culminated in the
classical work by \citet{Tholen84}. More recently, broadband
photometry has evolved in full-fledged spectroscopy using 
spectrographs equipped with CCDs. A commonly adopted taxonomic
classification based on spectra at visible wavelengths was published
by \citet{BusBin02}, and an extension to the near IR region was more
recently proposed by \citet{DeMetal09}.

In this paper we want to assess whether spectro-polarimetry may be
used to complement and refine the observing techniques of spectroscopy
and broadband polarimetry, that so far have been only {\it separately}
considered. For this reason, we have started a survey of
spectro-polarimetry of asteroids, to our knowledge the first of its
kind.

The taxonomic classifications of reflectance spectra by
\citet{Tholen84} and \citet{BusBin02} were based on Principal
Component Analysis of hundreds of objects. So far, our
spectro-polarimetric dataset is far too small to allow us any
systematic classification. This paper presents therefore the results
of a pilot project aimed at assessing the usefulness of further
investigations using this technique.
\vspace{-0.4cm}

\section{Observations}
\begin{table}
\input{Tab_Log}
\end{table}

\begin{figure}
\begin{center}
\scalebox{0.47}[0.50]{
\includegraphics*[1.4cm,1.7cm][19.9cm,26.1cm]{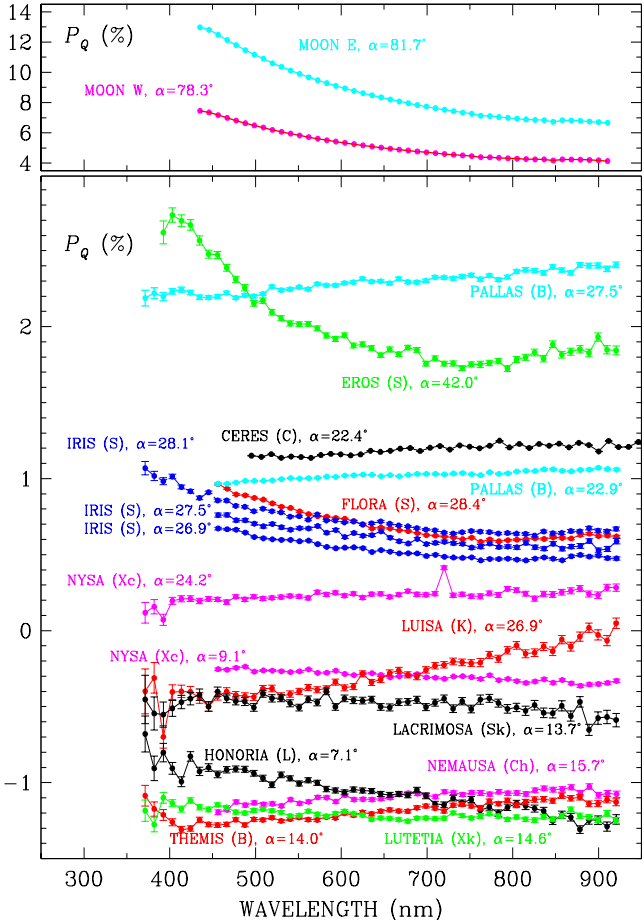}}
\end{center}
\caption{\label{Fig_Asteroids} Polarization spectra of 12 asteroids (bottom panel) 
and of two regions of the Moon (top panel).}
\end{figure}
We have obtained spectro-polarimetric measurements of a sample of
asteroids using the FORS2 instrument \citep{Appetal98} of the ESO Very
Large Telescope (VLT), and the ISIS instrument of the William Herschel
Telescope (WHT) of the Isaac Newton Group of Telescopes. During an
earlier VLT-FORS visitor mode run dedicated to the observations of the
Earthshine \citep{Steretal12} we have also observed the sunlit limb of
the Moon.

The instruments employed in our measurements are slit-fed and are
equipped with similar polarimetric optics, consisting of a retarder
waveplate and a beam-splitter polarizer: a Wollaston prism in case of
FORS2, and a Savart plate in case of ISIS.  The retarder waveplates
may be set at fixed position angles, allowing one to exploit the
advantages of the ``beam-swapping'' technique
\citep{Bagetal09}. Thanks to the beam-swapping technique, to the fact
that both instruments are slit-fed, and that the light reflected by
the target reaches the polarimetric optics without oblique
reflections, we were able to obtain very accurate measurements of the
continuum polarization. Observations with the FORS instrument were
obtained using grism 300V with and without order-sorting filter GG435,
covering the wavelength range 435--930\,nm and 390--930\,nm,
respectively. ISIS observations were obtained using grism R158R and
order-sorting filter GG495, covering the spectral range 480\,nm to
975\,nm.

Reductions of FORS data were performed with the aid of the ESO FORS
pipeline \citep{Izzetal10}, and dedicated FORTRAN routines. Spectra
obtained with ISIS were extracted then wavelength calibrated using
IRAF routines, and then combined with FORTRAN routines. Throughout
this paper we will refer to the reduced Stokes parameter $\pq(\lambda)
= Q/I$ representing the flux perpendicular to the plane
Sun-Object-Earth (the scattering plane) minus the flux parallel to
that plane, divided by the sum of the two fluxes. For symmetric
reasons, Stokes $U$ is expected to be zero. From the
spectro-polarimetric data we calculated synthetic BBLP values (see
Table~\ref{Tab_Log}).  Approximate reflectance spectra $\rf(\lambda)$
were obtained by dividing the intensity spectra by the spectrum of solar
analogue HD\,30246 observed on 2014-01-30, but without taking into
account wavelength dependent slit losses, and then normalised to $\lambda =
550$\,nm. Data were rebinned to a spectral bin of $\sim 11$\,nm.

Polarization spectra of our targets are shown in Fig.~\ref{Fig_Asteroids}. As
expected, we found positive polarization ({\it i.e.}, perpendicular to the
scattering plane) at phase-angles $\alpha \ga 20\degr$, and negative
polarization ({\it i.e.}, parallel to the scattering plane) at phase-angles
$\alpha \la 20\degr$. Remarkably, there is one exception: in spite of having
been observed at a phase-angle as large as $\sim 27\degr$, asteroid
(599)\,Luisa exhibits a negative polarization. This makes it a new
member of the class of the so-called Barbarians \citep{Celetal06},
{\it i.e.}, asteroids displaying an anomalous phase-polarization curve,
characterized by a very wide negative polarization branch, extending
up to $\alpha \sim 30$\degr.

\section{Discussion}
To discuss the diagnostic power of spectro-polarimetry, we are going to 
address the following inter-related questions.
\begin{itemize}
\item[{\it i)}]   Do polarization spectra depend on the phase-angle?
\item[{\it ii)}]  Do asteroids of a given taxonomic class have identical polarization spectra?
\item[{\it iii)}] Do asteroids of different taxonomic classes have different polarization spectra?
\item[{\it iv)}]  What is the relationship between polarization spectra and reflectance spectra?
\end{itemize}
Firm answers require observations of several asteroids per taxonomic
class with a homogeneous sampling of the phase-angle range. However,
even our limited dataset suggests some tentative answers, and,
most importantly, guides us on how to refine the strategy for future
observations.

We already know from classical BBLP measurements that the fraction of
linear polarization does depend on phase-angle. In this analysis, however,
we are more interested in the {\it shape} of the polarization
spectra. Observations of (2)\,Pallas and (7)\,Iris suggest that in the
positive branch, at least within limited phase-angle ranges, the shape
of the \pq\ spectra does not change, although observations of
(44)\,Nysa suggest that the shape of the \pq\ spectra obtained in the
positive branch may differ from that obtained in the negative branch.
We therefore introduce the polarization spectra normalised to the value
at $\lambda = 550$\,nm:
\[
\pqn(\lambda,\alpha)=\frac{\pq(\lambda,\alpha)}{\pq(\lambda=550\,{\rm nm},\alpha)} \; .
\]
The introduction of this new quantity allows us to compare data of
different objects obtained at different phase-angles, under the
approximation that, at least to first-order, the dependence of the
polarization upon phase-angle may be separated from the dependence
upon wavelength, in which case we have $\pqn(\lambda,\alpha) \simeq
\pqn(\lambda)$. We note that unless the \pq\ spectra cross the zero,
\pqn\ is always positive.

Answering questions {\it ii)} and {\it iii)} is equivalent to
explicitly addressing the issue of whether spectro-polarimetry brings
additional information than spectroscopy. Figure~\ref{Fig_Asteroids}
suggests that in the specific case of asteroids (7)\,Iris and (8)\,Flora
-- both S-class in the \citet{BusBin02} system, and both observed at
$\alpha \sim 28\degr$ -- the answer to question {\it ii)} is yes. To
address question {\it iii)} we may consider that asteroids (2)\,Pallas
(B-class), (7)\,Iris (S-class), and (599)\,Luisa (K-class) which were
all all observed close to $\alpha\sim 27\degr$, show rather different
\pq\ spectra. To proceed further, we can only compare observations of
different asteroids obtained at different phase-angles, making use of
the normalised polarization spectra \pqn\ introduced above.

\begin{figure}
\begin{center}
\scalebox{0.44}{
\includegraphics*[0.7cm,7.2cm][19.9cm,26.1cm]{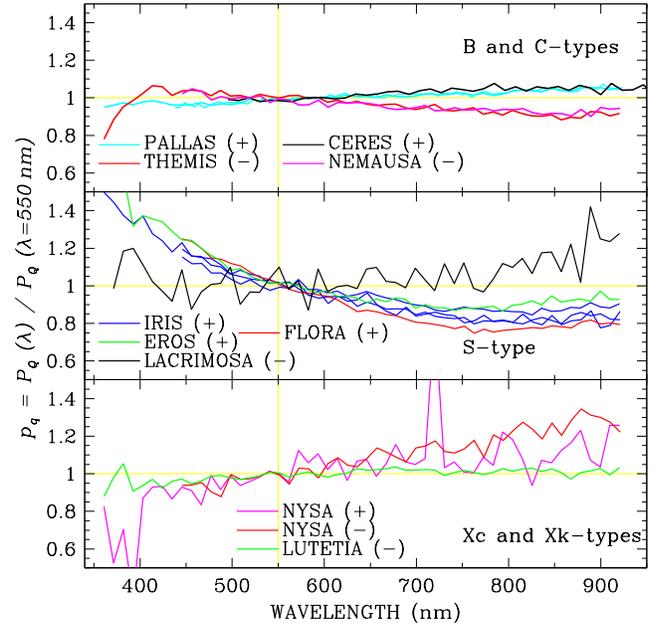}}
\end{center}
\caption{\label{Fig_Classes} \pqn\ spectra of asteroids
  ({\it i.e.}, \pq\ spectra normalised to
  the value at $\lambda=550$\,nm). The symbol (+) means that the
  spectrum was obtained in the positive branch, while the symbol (--)
  means that it was obtained in the negative branch.}
\end{figure}

The top panel of Fig.~\ref{Fig_Classes} shows the \pqn\ spectra of B-
and C-type asteroids. Asteroids (2)\,Pallas and (1)\,Ceres are both
observed in the positive branch, and share similar
\pqn\ spectra. Asteroids (21)\,Themis and (51)\,Nemausa are both
observed in the negative branch, and also share similar
\pqn\ spectra. In Fig.~\ref{Fig_Asteroids} we see that the
\pq\ spectra of B- and C-type asteroids always have a negative
wavelength gradient. We note that since in the negative branch the
gradients of \pqn\ and \pq\ spectra have opposite sign,
B- and C-type asteroids have d\pqn/d$\lambda <0$ in the
negative branch, and d\pqn/d$\lambda <0$ in the positive branch
(see Fig.~\ref{Fig_Classes}).

The mid panel shows the \pqn\ spectra of four S-type asteroids:
(7)\,Iris (observed three times around $\alpha \sim 28\degr$),
(433)\,Eros (a near-Earth asteroid observed at $\alpha=42.0\degr$),
(8)\,Flora (observed at $\alpha=28.4\degr$) and (208)\,Lacrimosa
(observed in the negative branch at $\alpha=13.7\degr$). The three
\pqn\ spectra of (7)\,Iris overlap each other well. The \pqn\ spectrum
of (433)\,Eros exhibits a marginally more pronounced concavity than
that of (8)\,Flora and (7)\,Iris, but we are not able to say whether
this (small) difference comes from the fact that Eros observations
were obtained at a quite different phase-angle ($\sim 14\degr$\ larger
than those of Flora and Iris), or because
we are observing objects with different surface structure. We note
that the \pq\ spectra obtained in the positive branch have a negative
gradient.  The \pq\ spectrum of (208) Lacrimosa, the only S-class
asteroid observed in the negative branch, also has a negative gradient
(which corresponds to a positive gradient for \pqn).  We therefore
conclude that the intermediate albedo S-class asteroids exhibit a
polarimetric behaviour opposite to that of low-albedo B- and C-class
asteroids, \emph{i.e.}, the gradient of their \pq\ spectra is always
negative.

The bottom panel of Fig.~\ref{Fig_Classes} shows that the
\pqn\ spectra of high-albedo Xc-class asteroid (44)\,Nysa are somewhat
similar both in the negative and in the positive branch, and similar
to the other X-class asteroid (21)\,Lutetia.  The slope of the
\pq\ spectra of (44)\,Nysa is negative in the negative branch, and
positive in the positive branch, therefore it must change its sign
somewhere around the inversion angle. We may speculate that this
feature is common to all high-albedo asteroids, but more data are
needed to confirm this.

\begin{figure}
\begin{center}
\scalebox{0.43}{
\includegraphics*[0.4cm,5.6cm][19.9cm,24.3cm]{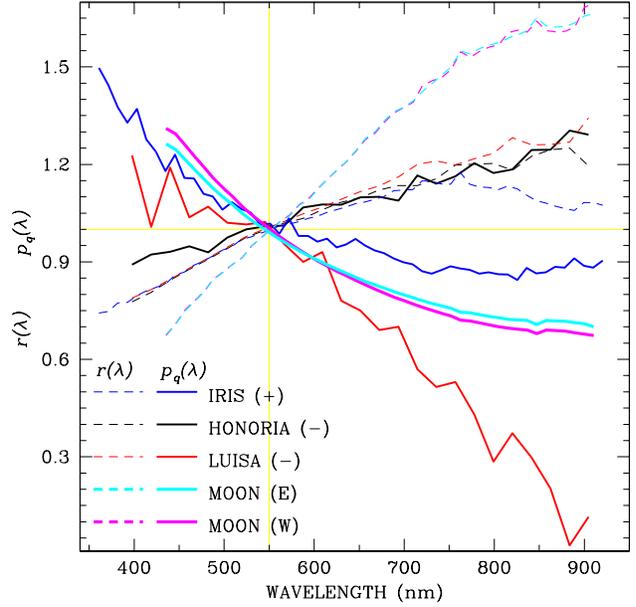}}
\end{center}
\caption{\label{Fig_Umov} Normalised polarization spectra 
\pqn\ (thick solid lines) and reflectance \rf\ spectra (thin
dashed lines) of two regions of the Moon and of three asteroids
of different taxonomic classes.
}
\end{figure}

We now consider the polarization spectra of two regions
at the limb of the Moon, one close to the Grimaldi crater, and one
close to the Mare Crisium, which are plotted in the top panel of
Fig.~\ref{Fig_Asteroids}.  Due to the high phase-angle value, both
lunar \pq\ spectra have a much higher amplitude than that observed for
asteroids. Compared among themselves, the two lunar spectra show a
similar trend but, due to the different phase-angle at which they were
obtained, have a quite different amplitude.  Once they are normalised,
they nearly overlap each other, as shown with the light blue and
magenta solid lines in Fig.~\ref{Fig_Umov}. 

Figure~\ref{Fig_Umov}  allows us
to compare spectropolarimetric data of asteroids of different
taxonomic classes and of the Moon, and also to make some
considerations about the reflectance spectra, that are shown with
dashed lines (normalised to $\lambda=550$\,nm).
First we consider the three asteroids (7)\,Iris, (236)\,Honoria, and
(599)\,Luisa, which, although belonging to different classes in the
\citet{BusBin02} system, were all classified as S-type in the Tholen
system. It is remarkable that, while their reflectance spectra appear
similar to each other (which explains their common classification in
the Tholen system), the \pqn\ spectra appear completely different from
each other! In particular, both (236)\,Honoria and (599)\,Luisa were
observed in the negative branch, but display polarization spectra with
opposite gradients: the \pqn\ spectrum of (599)\,Luisa (observed at
$\alpha \sim 27\degr$) has a strong negative gradient, i.e., the
absolute value of the polarization decreases with wavelength; {\it
  viceversa}, the absolute value of the polarization of (236)\,Honoria
(observed at $\alpha \sim 7\degr$) strongly increases with wavelength.
The \pqn\ spectra of the Moon are steeper than that of S-type
asteroid (7) Iris (blue solid lines), but less steep than K-type
asteroid (599)\,Luisa (red solid line). It is likely that the
difference in slope are determined by a remarkable diversity of the
surface structures and compositions.

In general, we expect that higher albedo
corresponds to smaller polarization, and lower albedo to higher
polarization. Indeed the behaviour of lunar spectra confirms 
the results by \citet{Doletal71} that for lunar regions, the
polarization and reflectance spectra obey to the
Umov law \citep{Umov05}, i.e., $\pq(\lambda) \propto 1/\rf(\lambda)$.
The Umov law is also valid for asteroids (7)\,Iris and (599) Luisa,
but in the case of asteroid (236)\,Honoria, both the absolute value of the
polarization and the reflectance have a positive gradient, {\it i.e.}
both polarization and albedo increase with wavelength. A similar
behaviour is exhibited by (21)\,Lutetia (not shown in the Figure),
that in the Tholen system had been classified as M-type.  This is
another aspect of the phenomenon discussed by \citet{Beletal09} who
discovered that in M-type and S-type asteroids, which have higher
albedo in the red than in the blue, the minimum of the polarization
curves is deeper in the red than in the blue. Umov law is rooted on
the basic mechanism described by the Fresnel laws. Perhaps it is not
surprising that it is violated in those conditions when Fresnel laws
cannot even explain the orientation of the polarization. This
phenomenon deserves further observational and theoretical
investigation. In particular it would be interesting to assess if it
manifests itself only at small phase-angles, when the polarization is
parallel to the scattering plane (being perhaps linked to the coherent
backscattering mechanism), or if it may be observed also at large
phase-angles.

(236)\,Honoria is a known Barbarian.  Our discovery that also
(599)\,Luisa is a Barbarian is particularly interesting. In the space
of orbital proper elements this asteroid is located in a
high-inclination region where other Barbarians are also
present, {\it i.e.}, (387)\,Aquitania, (980)\,Anacostia and
(729) Watsonia. The latter is the lowest-numbered member of a
dynamical family \citep{Bojan11,Miletal14} which has been
found by \citet{Celetal14} to be a reservoir of small
Barbarians. Moreover, spectroscopic data show that
(387)\,Aquitania, (980)\,Anacostia and (599)\,Luisa
have peculiarly high abundances of the spinel mineral, up to 30\,\%
\citep{Sunetal08}. The link between the Barbarian polarimetric
behaviour and a composition rich in spinel is therefore further
confirmed by our discovery that the spinel-rich asteroid (599)\,Luisa
is also a Barbarian. We remind that (236)\,Honoria
and (599)\,Luisa display opposite polarimetric gradients (see Fig~\ref{Fig_Umov}).
Perhaps these differences are due to the large gap
in phase-angle at which the observations were
taken (though both in the negative branch), which would
imply that the wavelength gradient of the polarization changes
its sign in the negative branch. If instead the difference of the polarization
spectra reflects a difference in structure surface and composition,
we may have found a hint to the existence of different categories of
Barbarians. More data are needed to confirm this.

\section{Conclusions}
We have obtained a number of polarization spectra of asteroids and the
Moon. From their analysis we {\it tentatively} suggest that
\pq\ spectra of low albedo asteroids always have a positive gradient,
and intermediate albedo asteroids always have a negative gradient:
this would be a confirmation of preliminary results obtained in the
pioneering works by \citet{LupKis95} and \citet{Beletal09}, based on
multi-colour BBLP data. Polarization spectra of high-albedo asteroid
(44)\,Nysa has a positive gradient in the positive branch, and a
negative gradient in the negative branch, but more observations
are needed to check if this result can be generalized to a wide class
of asteroids.

We have found strong evidence that the Umov law may be violated:
observed in the negative branch, both reflectance and polarization of
asteroid (236)\,Honoria strongly increase with wavelength. We have
also discovered that (599)\,Luisa is a member of the Barbarian class
of asteroids.

We have shown that two objects belonging to the S-class observed at
the same phase-angle have nearly identical polarization spectra, but
we have also found that three objects, (7)\,Iris, (236)\,Honoria and
(599)\,Luisa, have relatively similar optical reflectance spectra but
totally different polarizations spectra. Particularly puzzling is the
difference between the polarization spectra of Honoria and Luisa,
which are both members of the Barbarian class of asteroids. We do not
know if this diversity is a consequence of the fact that these objects
were observed at a different phase-angle, or if it originates from a
remarkably different surface structure.

In asteroid spectroscopy, the choice of the solar analogue used for
the normalisation of the intensity spectra, and the quality of the
calibration of the atmospheric extinction play a crucial role on the
final data quality, and, ultimately, on the spectral classification of
asteroids. By contrast, spectro-polarimetric measurements are robust,
nearly independent of atmospheric conditions, and do not require any
calibration with a solar analogue star. Provided that instrumental
polarization is low and under control, they may be perfectly
reproduced even with different instruments. Spectro-polarimetric
techniques still allow us to simultaneously obtain reflectance
spectra, provided that the usual calibrations are performed.

We suggest that spectro-polarimetric analysis of asteroids should
complement traditional spectroscopic measurements and
classification. In the longer term, any physical model capable of
reproducing the observed reflectance spectra should also be tested
against its capability to reproduce the observed spectro-polarimetric
data.

\section*{Acknowledgments}
SB and AC gratefully acknowledge support from COST Action MP1104 
``Polarimetry as a tool to study the solar system and beyond''.
Observations were performed with ING
Telescopes under programme W/2014A/5 and with ESO Telescopes at the
La Silla-Paranal Observatory under programme IDs 087.C-0040(A) and 092.C-0639.
We thank WHT staff, the ESO Paranal SCIOPS Team and User
Support Division for their excellent support to the observations.

\label{lastpage}

\end{document}

%% file: Tab_Log.tex
\caption{\label{Tab_Log}
BBLP values in the Bessel $VRI$ filters from \pq\ spectra. Photon-noise
is negligible, and accuracy is limited by 
instrumental polarization, which we estimate $\le 0.1$\,\%.  
The double taxonomy classification given in col.~2 are from
\citet{Tholen84} (left) and \citet{BusBin02} (right). Asteroid
observations were obtained from September 2013 to March 2014. (1)
Ceres was observed with ISIS, all the remaining targets with FORS.
The Moon was observed with FORS in April and June 2011.
}
\begin{small}
\begin{center}
\begin{tabular}{r@{\,}llrrrr}
\hline\hline
\multicolumn{2}{c}{Object}       &
\multicolumn{1}{c}{Class}        &
\multicolumn{1}{c}{$\alpha$}     &
\multicolumn{1}{c}{$ V $}        &
\multicolumn{1}{c}{$ R $}        &
\multicolumn{1}{c}{$ I $}       \\
\multicolumn{2}{c}{}             &
\multicolumn{1}{c}{}             &
\multicolumn{1}{c}{}             &
\multicolumn{1}{c}{$(\%)$}       &
\multicolumn{1}{c}{$(\%)$}       &
\multicolumn{1}{c}{$(\%)$}      \\
\hline
  (1) &Ceres   & G/C  &$22.4\degr$&$ 1.17$&$ 1.21$&$ 1.25$\\ [2mm]
  (2) &Pallas  & B/B  &$27.5\degr$&$ 2.25$&$ 2.29$&$ 2.33$\\ 
      &        &      &$22.9\degr$&$ 0.99$&$ 1.00$&$ 1.03$\\ [2mm]
  (7) &Iris    & S/S  &$26.9\degr$&$ 0.58$&$ 0.52$&$ 0.48$\\
      &        &      &$27.5\degr$&$ 0.68$&$ 0.62$&$ 0.56$\\
      &        &      &$28.2\degr$&$ 0.75$&$ 0.68$&$ 0.64$\\ [2mm]
  (8) &Flora   & S/S  &$28.4\degr$&$ 0.78$&$ 0.68$&$ 0.60$\\ [2mm]
 (21) &Lutetia & M/Xk &$14.6\degr$&$-1.19$&$-1.23$&$-1.23$\\ [2mm]
 (24) &Themis  & C/B  &$14.0\degr$&$-1.23$&$-1.18$&$-1.12$\\ [2mm]
 (44) &Nysa    & E/Xc &$ 9.1\degr$&$-0.27$&$-0.30$&$-0.32$\\ 
      &        &      &$24.2\degr$&$ 0.23$&$ 0.24$&$ 0.25$\\ [2mm]
 (51) &Nemausa &CU/Ch &$15.7\degr$&$-1.11$&$-1.10$&$-1.06$\\ [2mm]
(208) &Lacrimosa&S/Sk &$13.7\degr$&$-0.46$&$-0.47$&$-0.50$\\ [2mm]
(236) &Honoria & S/L  &$ 7.1\degr$&$-1.00$&$-1.08$&$-1.17$\\ [2mm]
(433) &Eros    & S/S  &$42.0\degr$&$ 1.99$&$ 1.87$&$ 1.86$\\
(599) &Luisa   & S/K  &$26.9\degr$&$-0.39$&$-0.30$&$-0.16$\\ [2mm]
      &Moon E  & n.a. &$81.7\degr$&$ 9.86$&$ 8.28$&$ 7.07$\\ 
      &Moon M  & n.a. &$78.3\degr$&$ 5.81$&$ 4.99$&$ 4.36$\\ 
\hline
\end{tabular}
\end{center}
\end{small}